\documentstyle[12pt]{article}
\def\be{\begin{equation}}
\def\ee{\end{equation}}
\def\ba{\begin{array}}
\def\ea{\end{array}}

\def\ni{\noindent}
\def\bb{\begin{thebibliography}{99}\small\addtolength{\itemsep}{-6pt}}
\def\eb{\end{thebibliography}}
\newcommand{\dsp}{\displaystyle}
\newcommand{\scs}[1]{{\scriptscriptstyle #1}}
\newcommand{\tr}{{\rm tr}\,}
\newcommand{\hc}{{\rm h.c.}\,}
\newcommand{\eqr}[1]{(\ref{#1})}
\renewcommand{\d}{\partial}

\newcommand{\N}{\scs N}
\def\cS{{\cal S}}
\def\a{\alpha}
\def\b{\beta}
\def\g{\gamma}
\def\e{\;{\rm e}}
\def\l{\lambda}
\def\L{\Lambda}
\def\d{\partial}
\def\s{\sigma}
\def\D{\Delta}
\def\O{\Omega}
\def\gq{\frac{\g^q}{q!}}
\def\ln{\frac{\l_o}{2N}}
\def\nl{\frac{N}{\l_o}}
\def\lo{\l_o}
\def\da{\dagger}
\def\dc{\delta c}
\def\eps{\epsilon}
\def\det{\;\hbox{det}}
\def\dpk{\frac{\d}{\d\phi_k}\Big|_{u=I}}
\def\taa{\tau_{\a\a}^r(u)}
\def\tauo{\tau_{\a\a}^r(\O^\da W\O)}

\begin{document}\begin{titlepage}
\begin{flushright}TAUP-2204-94\\September 1994\end{flushright}
\begin{center}\vskip1cm
{\LARGE\bf Lattice QCD as a theory of}\\ \vskip.2cm
{\LARGE\bf interacting surfaces}\\ \vskip1.5cm
{\large\bf Boris Rusakov} \\ \vskip1cm
{\em High Energy Section, ICTP,\\P.O.Box 586, I-34014 Trieste, Italy\\
       and\\ School of Physics and Astronomy \\
       Tel-Aviv University, Tel-Aviv 69978, Israel} \\ \vskip2.5cm
{\large\bf Abstract.}\end{center}

Pure gauge lattice QCD at arbitrary $D$ is considered. Exact integration
over link variables in an arbitrary $D$-volume leads naturally to an
appearance of a set of surfaces filling the volume and gives an exact
expression for functional of their boundaries. The interaction between
each two surfaces is proportional to their common area and is realized
by a non-local matrix differential operator acting on their boundaries.
The surface self-interaction is given by the QCD$_2$ functional of
boundary. Partition functions and observables (Wilson loop averages) are
written as an averages over all configurations of an integer-valued field
living on a surfaces.

\end{titlepage}
\newpage

\section{Introduction.}

Pure gauge QCD (quantum gluodynamics) in $D$ dimensions has been
reformulated by K.Wilson in terms of collective (lattice) variables
\cite{Wils}:\be S=\nl\sum_f\tr(U_f+U_f^\da) \label{wact}\ee
where sum goes over all faces $f$ of $D$-dimensional lattice,
$U_f=\prod_{i\in f}U_i$ with $U_i$ being unitary matrix (U(N) or SU(N))
attached to $i$-th link of a lattice, $\lo$ is the bar (lattice)
coupling constant. In continuum limit, at any $D\leq 4$, $\lo$ goes to
zero. At $D=2$ $~\lo\sim\eps^2$, at $D=3$ $~\lo\sim\eps$ and at $D=4$
$~\lo\sim-\frac{1}{\log\eps}$ where $\eps$ is the linear size of a
lattice.

The partition function is defined as an integral over all link variables,
\be Z=\int\prod_i dU_i\e^S, \ee while the simplest observable, the Wilson
loop average, is
\be W(C)=\frac{1}{NZ}\int\prod_i dU_i\e^S \tr\Big(\prod_{j\in C}U_j\Big)
\;\;, \label{W}\ee where $C$ is a closed contour.

In order to integrate exactly over unitary matrices one makes the group
Fourier transformation of the face variables, first manifestly used in
$D=2$ by A.A.Migdal \cite{Mig}: \be\e^{\nl\tr(U_f+U_f^\da)}=\sum_{r}
d_r\L_r\Big(\nl\Big)\chi_r(U_f)\;\;,\label{exp}\ee \be d_r\L_r\Big(\nl
\Big)=\int dU\chi_r(U)\e^{\nl\tr(U+U^\da)}\;\;,\label{coef}\ee
where $r$ is irreducible representation of the gauge group, $\chi_r(U)$
is its character and $d_r=\chi_r(I)$ is its dimension,
\be d_r = \prod_{1\leq i < j \leq N} \Big(1+\frac{n_i-n_j}{j-i}\Big)
\;\;.\label{dim}\ee Here and below we use the standard parametrization of
$r$ by its highest weight components, $n_1\ge ...\ge n_\N$, associated
with a lengths of lines in the Young table.

Direct calculation of \eqr{coef} gives\be\L_r\Big(\nl\Big)=\frac{1}{d_r}
\det_{ij}I_{n_i-i+j}\Big(\frac{2N}{\lo}\Big)\;\;,\label{la}\ee
where $I_n(x)$ is the modified Bessel function.

In $D=2$, an exact solution of the model can be obtained using only
orthogonality of characters, \be\int dU\chi_{r_1}(AU)\chi_{r_2}(U^\da B)=
\delta_{r_1,r_2}\frac{\chi_{r_1}(AB)}{d_{r_1}}\;\;.\label{ort}\ee
The result of integration inside a disk of area $A$ (therefore, its
lattice area is $A/\eps^2$) gives {\it functional of boundary}
\footnote{By ``boundary", here and below, we equally imply the
geometrical boundary and the product of unitary matrices attached to it.}
$\Gamma$ \cite{Mig}:\be Z_{\rm latt}(\Gamma)=
\sum_{r}d_r\L_r^{A/\eps^2}\chi_r(\Gamma)\;.\label{fblat}\ee
In the continuum limit, $\eps\to 0$ and $\lo=\l\eps^2$, we need an
asymptotic expansion of $\L_r$, which can be found from \eqr{coef} by the
saddle point method to give: \be\L_r\sim 1-\frac{\lo C_2(r)}{2N}+O(\lo^2)
\label{asym}\ee up to representation-independent factor. Here, $C_2(r)$
is the eigenvalue of the second Casimir operator,
\be C_2(r) = \sum_{i=1}^{N} n_i(n_i +N+1-2i)\;\;.\label{Casimir}\ee
Thus, the continuum limit of \eqr{fblat} is defined by the substitution
$$\L_r^{A/\eps^2}\to\exp\Big(-\frac{\l A}{2N}C_2(r)\Big)$$ and takes
the form\be Z(\Gamma)=\sum_{r}d_r\exp\Big(-\frac{A}{2N}C_2(r)\Big)
\chi_r(\Gamma)\;.\label{fb1}\ee The same can be done in the case of
non-trivial topology. The results are \cite{Rus}
\footnote{See also \cite{Witten}.}:

\ni (i) functional of boundaries $\Gamma_i$, $i=1,...,n$ of a sphere with
$n$ holes and of area $A$ (continuum coupling constant $\l$ is absorbed
into the area):\be Z(\Gamma_1,...,\Gamma_n)=\sum_{r}d_r^{2-n}
\exp\Big(-\frac{A}{2N}C_2(r)\Big)
\prod_{i=1}^n\chi_r(\Gamma_i)\;,\label{fb}\ee

\ni (ii) partition function for a closed surface of a genus $g$
and of area $A$:\be Z_g(A)=\sum_{r} d_r^{2(1-g)}
\exp\Big(-\frac{A}{2N}C_2(r)\Big)\;\;,\label{Ztwo}\ee

\ni (iii) Wilson loop average:
\be W_g(C)=\sum_{r_1,...,r_m}\Phi_{r_1...r_m}\prod_{k=1}^m
d_{r_k}^{2(1-g_k)}\exp\Big(-\frac{A_k}{2N}C_2(r_k)\Big)\;\;,
\label{Wtwo}\ee where $m$ is the number of windows, $A_k$ is the area of
a window, $g_k$ is the ``genus per window" and coefficient
$\Phi_{r_1...r_m}$ is the U(N) (SU(N)) group factor dependent on the
contour topology (see Ref.\cite{Rus} for details).
\bigskip

In $D>2$, the orthogonality condition \eqr{ort} is not enough
to perform an integration over link variables since there are more
than two plaquettes match at each link. Formally, we still could
integrate using known formulas for tensor product of irreducible
representations. This results in a sums over internal spaces of
representations. The Clebsch-Gordan coefficients entering in these
expansions are not known in any compact and general form. Besides, after
integration we should perform a summation of a resulting expressions
weighted with these Clebsch-Gordan coefficients, which makes the problem
to be extremely difficult. However, the problem becomes less hopeless if
one guesses that the three steps, namely, expansion into representations,
integration over link variables and then back re-summation, could be
performed in one step if some adequate variables are found, which should
be a combination of one-link integral and of a sum over representations.

In the next two sections we present such a variables and describe the
procedure of integration in an arbitrary (lattice) $D$-volume.

\section{One-link integral.}

Though our final results will not depend on choice of a lattice, we
start for concreteness from the quadrilated regular lattice.

\begin{figure}\centering\begin{picture}(200,100)(- 100,- 50)
\thicklines
\put(-90, 35){\line(-2,1){30}}\put(-90, 35){\line(0,-1){9}}
\put(-120, 50){\line(0,-1){39}}\put(-120, 30){\vector(0,1){4}}
\put(-70, 35){\line( 2,1){30}}\put(-70, 35){\line(0,-1){9}}
\put(-40, 50){\line(0,-1){39}}\put(-40, 30){\vector(0,1){4}}
\put(-74,27){\line(2,-1){40}}\put(-74,27){\line(0,-1){40}}
\put(-74,-13){\line(2,-1){40}}\put(-34, 7){\line(0,-1){40}}
\put(-34,-13){\vector(0,1){4}}\put(-74, 7){\vector(0,-1){4}}
\put(-86,27){\line(-2,-1){40}}\put(-86,27){\line(0,-1){40}}
\put(-86,-13){\line(-2,-1){40}}\put(-126, 7){\line(0,-1){40}}
\put(-86, 7){\vector(0,-1){4}}\put(-126,-13){\vector(0,1){4}}
\put(-138,-13){\makebox(0,0){$W_1$}}\put(-132,30){\makebox(0,0){$W_2$}}
\put(-28,30){\makebox(0,0){$W_3$}}\put(-22,-13){\makebox(0,0){$W_4$}}
\put(-94,7){\makebox(0,0){$U$}}\put(-63,7){\makebox(0,0){$U$}}
\put(80, 30){\line(-2,1){40}}\put(40, 50){\line(0,-1){39}}
\put(40, 30){\vector(0,1){4}}\put(90, 35){\line( 2,1){30}}
\put(90, 35){\line(0,-1){9}}\put(120, 50){\line(0,-1){29}}
\put(120, 30){\vector(0,1){4}}\put(90, 35){\line(2,-1){40}}
\put(80,30){\line(2,-1){40}}\put(70,25){\line(2,-1){40}}
\put(70,-15){\line(2,-1){40}}\put(70,25){\line(-2,-1){40}}
\put(70,-15){\line(-2,-1){40}}
\put(30,5){\line(0,-1){40}}\put(110,5){\line(0,-1){40}}
\put(120,10){\line(0,-1){40}}\put(120,-30){\line(-2,1){10}}
\put(130,15){\line(0,-1){40}}\put(130,-25){\line(-2,1){10}}
\put(30,-15){\vector(0,1){4}}\put(110,-15){\vector(0,-1){4}}
\put(120,-10){\vector(0,-1){4}}\put(130,-5){\vector(0,-1){4}}
\put(70,-32){\makebox(0,0){$W_1W_4^\da$}}
\put(140,30){\makebox(0,0){$W_3W_4^\da$}}
\put(20,30){\makebox(0,0){$W_2W_4^\da$}}
\thinlines\put(- 15,10){\vector(1,0){30}}
\end{picture}
\caption[x]{\footnotesize Gluing in three-dimensions.}
\label{fig1}\end{figure}
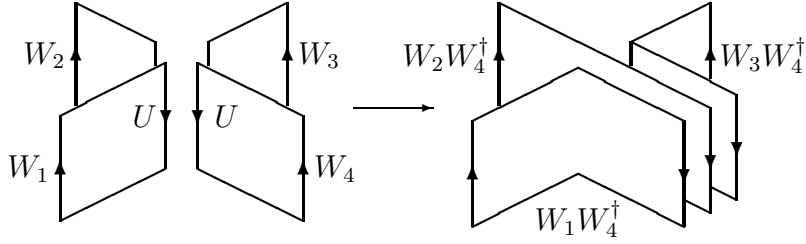

There are $2D-2$ plaquettes interacting through each link on a
quadrilated $D$-dimensional lattice. The one-link integral has the form
\be\int dU\prod_{k=1}^{2D-2}\e^{\nl\tr W_kU+\hc}\;,\label{wi}\ee
where by $W_k$ we denote a product of three other unitary matrices
in a $k$-th plaquette (see left side of Fig.\ref{fig1} for $3D$ example).
In the heat kernel framework, \eqr{wi} is equivalent to
\be \sum_{r_1...r_{2D-2}}\prod_{k=1}^{2D-2}d_{r_k}\e^{-\ln C_2(r_k)}
\int dU\prod_{k=1}^{2D-2}\chi_{r_k}(W_kU)\;.\label{hk}\ee
Expressions \eqr{wi} and \eqr{hk} are identical up to the $O(\lo)$ order.
The difference appears in order $O(\lo^2)$.

Making shift $U\to W_j^\da U$ (the Haar measure $dU$ is invariant with
respect to such transformations) we see that the integral \eqr{wi},
\eqr{hk} depends on $2D-1$ boundaries
$W_kW_j^\da$, $j\neq k$. The shift can be made in $2D-2$ possible
directions corresponding to $2D-2$ possible choices of $j$. The picture
corresponding to one of a four possible directions of gluing in $D=3$ is
presented on Fig.\ref{fig1}.

Thus, after one-link integration we obtain $2D-1$ two-plaquette
surfaces with some interaction between them. We are going now to
calculate this interaction and to continue the procedure consequently
for all links.

Let us start from the heat kernel representation \eqr{hk} and consider
for simplicity the case of only three plaquettes,
\be \sum_{r_1,r_2,r_3}d_{r_1}d_{r_2}d_{r_3}
\e^{-\ln(C_2(r_1)+C_2(r_2)+C_2(r_3))}
\int dU\chi_{r_1}(W_1U)\chi_{r_2}(W_2U)\chi_{r_3}(U)\;,\label{i1}\ee
where $W_1$ and $W_2$ are the boundaries of the two-plaquette surfaces.

We are going now to calculate exactly the quantity
\be \sum_rd_r\e^{-\ln C_2(r)}\int dU
\chi_{r_1}(W_1U)\chi_{r_2}(W_2U)\chi_r(U)\;.\label{int}\ee

A direct strategy would be to integrate first over $U$ and then to
take the sum over $r$ and over its internal sub-space. As we have
mentioned in Introduction this way is extremely difficult. Instead of
that, we first replace the sum over $r$ by the original Wilson term
$\exp\nl\tr(U+U^\da)$ and then derive the heat kernel exponent as a first
order non-zero term in $\lo$-expansion of an integral over $U$. This will
give an adequate variables for \eqr{int}.

To derive $\lo$-expansion of the integral
\be\frac{1}{f(\lo)}\int dU\e^{\nl\tr(U+U^\da)}
\chi_{r_1}(W_1U)\chi_{r_2}(W_2U)\;\;,\hskip.5cm
f(\lo)=\int dU\e^{\nl\tr(U+U^\da)}\;\;,\label{a1}\ee
we diagonalize $U$, $U=\O u\O^\da$ where
$u=diag\{\e^{i\phi_1},...,\e^{i\phi_\N}\}$, and integrate over diagonal
$u$ near the saddle point $u=I$. Then, \eqr{a1} takes the form:
\be\ba{lll}&\dsp\chi_{r_1}(W_1)\chi_{r_2}(W_2)\Big(1-\ln
\Big[C_2(r_1)+C_2(r_2)-2C_1(r_1)C_1(r_2)\Big]\Big)\\&\\
&\dsp +\lo\frac{N+1}{N}\sum_{k=1}^N\int d\O
\dpk\chi_{r_1}(W_1\O u\O^\da)\dpk\chi_{r_2}(W_2\O u\O^\da)
\;+O(\lo^2)\;,\ea\label{a2}\ee where the relation
\be\sum_{k=1}^N\dpk\chi_r(W\O u\O^\da)=iC_1(r)\chi_r(W)\;,
\hskip1.5cm C_1(r)=\sum_{k=1}^Nn_k\label{der}\ee has been used
($C_1(r)$ is the first Casimir operator eigenvalue). Indeed,
\be\sum_{k=1}^N\dpk\chi_r(W\O u\O^\da)=\sum_{\a}\tauo\sum_{k=1}^N
\dpk\taa\;,\label{indeed}\ee where $\tau_{\a\b}^r$ is the matrix element
of an $r$-th irreducible representation. There are only diagonal matrix
elements in \eqr{indeed} since $u$ is diagonal. The matrix label $\a$
is parametrized by the following Gelfand-Zetlin patterns \cite{Zel}:
\be\a=\left(\ba{ccccc}
&&n_1^\N~~~~~~~n_2^\N~~~~~~~n_3^\N~~~~~~~\dots~~~~~~~n_\N^\N&& \\
&&n_1^{\N-1}~~~~n_2^{\N-1}~~~~\dots~~~~n_{\N-1}^{\N-1}&& \\
&&\vdots~~~~~~\vdots~~~~~~\vdots&& \\ & &n_1^2~~~~~~n_2^2& &  \\
& &n_1^1& &  \ea \right) \;, \label{GZ}\ee where
\be n_k^j\ge n_k^{j-1}\ge n_{k+1}^j\;,\ee (the top level numbers are
coincide with the highest weight components: $n_k^\N=n_k, ~~~k=1,...,N$).
Then\be\taa =\prod_{j=1}^{N}\e^{i\phi_j\dc_j},\ee where $c_j$ is a sum of
numbers of $j$-th level,\be c_j=\sum_{k=1}^{j}n_k^j\;\;,\hskip1cm c_0=0
\label{c}\ee and $\dc_j=c_j-c_{j-1}$. Besides,
\be\sum_{k=1}^N\dc_k=c_\N\equiv C_1(r)\;.\ee
Then, the relation \eqr{der} becomes obvious.

To make further calculation we use the formula ({\it Weyl second formula
for characters}): \be\chi_r(W)=\det_{ij}\s_{n_i-i+j}(W)\;,\label{weyl}
\ee where numbers $n_i$'s are the highest weight components of $r$ and
$\s_n(W)$ is the character of representation $\{n,0,...,0\}$
\footnote{In formula \eqr{weyl}, $\s_n$ with $n<0$ might occur. The
convention for that case is $\s_n=0$.}. This character can be written as
\be\s_n(W)=w_{n,1}(\tr W)^n+\sum_{q=2}^nw_{n,q}(\tr W)^{n-q}\tr W^q
\label{sig}\ee (an exact form of a coefficients $w_{n,i}$ is not needed
for the following consideration). Since
\be -i\dpk\s_n(W\O u\O^\da)=(\O^\da W\O)_{kk}\s_{n-1}(W)
+\sum_{q=2}^nqw_q(\O^\da W^q\O)_{kk}(\tr W)^{n-q} \ee
and \be\sum_{k=1}^N\int d\O(\O^\da W_1\O)_{kk}(\O^\da W_2\O)_{kk}
=\frac{1}{N+1}\Big(\tr W_1~\tr W_2+\tr W_1W_2\Big)\;\;,\ee
we find:\be\ba{lll}&\dsp(N+1)\sum_{k=1}^N\int d\O\dpk\s_n(W_1\O u\O^\da)
\dpk\s_m(W_2\O u\O^\da)\\&\\&\dsp =\Big(\tr\d_{W_1} ~\tr\d_{W_2}
+\tr\d_{W_1}\d_{W_2}\Big) \s_n(W_1)\s_m(W_2)\;\;,\ea\label{forsig}\ee
where matrix elements of $\d_W$ are defined by\be\Big(\d_W\Big)_{jk}=
i\sum_{n=1}^NW_{jn}\frac{\d}{\d W_{kn}}\;,\hskip1.5cm
\frac{\d}{\d W_{kn}}W_{jm}=\delta_{jk}\delta_{mn}\;,\label{dw}\ee
i.e., matrix elements $W_{jk}$ of fundamental representation has to be
considered as an independent variables (in other words, an action of
derivative \eqr{dw} is defined on the GL(N) group). An important property
of derivative $\d_W$ is its invariance under right group transformations
\footnote{We could define $\d_W$ as
$$\Big(\d_W\Big)_{jk}=i\sum_{n=1}^NW_{nj}\frac{\d}{\d W_{nk}}\;.$$
Then, $\d_W$ is invariant under the left shift $W\to VW$.},
\be\d_W=\d_{WV}\;,\hskip1.5cm V\in{\rm GL(N)}\;.\label{shift}\ee
It is not difficult now to generalize \eqr{forsig} to the case of
arbitrary representations: \be\ba{lll}&\dsp (N+1)\sum_{k=1}^N\int d\O
\dpk\chi_{r_1}(W_1\O u\O^\da)\dpk\chi_{r_2}(W_2\O u\O^\da)\\&\\
&\dsp =\Big(\tr\d_{W_1} ~\tr\d_{W_2}+\tr\d_{W_1}\d_{W_2}\Big)
\chi_{r_1}(W_1)\chi_{r_2}(W_2)\;\;.\ea\ee Finally, since
\be\tr\d_W\chi_r(W)=iC_1(r)\chi_r(W)\ee (compare with \eqr{der}),
equation \eqr{a1} takes the form \footnote{In what follows we put
$f(\lo)=1$ since this factor is representation-independent and,
therefore, precisely cancels in any physical quantity (normalized to
partition function).}:
\be\ba{lll}&\dsp\int dU\e^{\nl\tr(U+U^\da)}
\chi_{r_1}(W_1U)\chi_{r_2}(W_2U)\\&\\&=\Big(1-\ln
\Big[C_2(r_1)+C_2(r_2)-2\tr\d_{W_1}\d_{W_2}\Big]\Big)
\chi_{r_1}(W_1)\chi_{r_2}(W_2)\;+O(\lo^2)\;.\ea\ee
Therefore, \be\ba{lll}\dsp\sum_rd_r\e^{-\ln C_2(r)}
&\dsp\int dU\chi_{r_1}(W_1U)\chi_{r_2}(W_2U)\chi_r(U)\\&\\&\dsp =
\e^{-\ln(C_2(r_1)+C_2(r_2)-2\tr\d_{W_1}\d_{W_2})}
\chi_{r_1}(W_1)\chi_{r_2}(W_2)\;.\ea\label{integral}\ee

As explained above, the result \eqr{integral} is {\it exact}, i.e.,
valid for arbitrary $\lo$. It can be also viewed as follows. Equation
\eqr{int} can be written as \be\ba{lll}\dsp\sum_r d_r\e^{\ln\D_{W=I}}
&\dsp\int dU\chi_{r_1}(W_1U)\chi_{r_2}(W_2U)\chi_r(WU)\\&\\
&\dsp =\sum_{k=1}^\infty\frac{\l_o^k}{k!(2N)^k}\D_{W=I}^k\int dU
\chi_{r_1}(W_1U)\chi_{r_2}(W_2U)\sum_r d_r\chi_r(WU)\;,\ea\label{i2}\ee
where $\D_W$ is an invariant Laplace-Beltrami operator which is the
differential operator with respect to parameters of $W$. A character
of an arbitrary irreducible representation is the eigenfunction of
$\Delta$, \be\D_W\chi_r(W)=-C_2(r)\chi_r(W)\;.\ee
Using the completeness condition for a characters,
\be \sum_r d_r\chi_r(W)=\delta(W,I)\;,\ee we write \eqr{i2} as
\be\sum_{k=1}^\infty\frac{\l_o^k}{k!(2N)^k}\D_{W=I}^k~
\chi_{r_1}(W_1W)\chi_{r_2}(W_2W)\;.\label{i3}\ee Actually,
we have checked above that \be\D_W\chi_{r_1}(W_1W)\chi_{r_2}(W_2W)=
-(C_2(r_1)+C_2(r_2)-2\tr\d_{W_1}\d_{W_2})\chi_{r_1}(W_1W)\chi_{r_2}(W_2W)
\;.\ee Therefore,\be\D_W^k\chi_{r_1}(W_1W)\chi_{r_2}(W_2W)=
(-)^k(C_2(r_1)+C_2(r_2)-2\tr\d_{W_1}\d_{W_2})^k\chi_{r_1}(W_1W)
\chi_{r_2}(W_2W)\;,\ee which confirms \eqr{integral}.

A generalization to the case of arbitrary number $P$ of plaquettes
joining at one link is straightforward: \be\ba{lll}&\dsp\sum_rd_r\e^{-\ln
C_2(r)}\int dU\chi_r(U) \prod_{p=1}^P\chi_{r_p}(W_pU)\\&\\&\dsp =
\exp -\ln\Big(\sum_{p=1}^PC_2(r_p)-2\sum_{<pq>}\tr\d_{W_p}\d_{W_q}\Big)
\prod_{p=1}^P\chi_{r_p}(W_p)\;.\ea\label{integral_gen}\ee

Operator $\exp\frac{\l_o}{N}\tr\d_{W_i}\d_{W_j}$ acts on a characters
in a simple way. In Appendix we give several examples corresponding to
several low-dimensional representations. It is worth
mentioning, that although the formula \eqr{weyl} is written for
non-negative signatures ($n_N\geq 0$) only, the results \eqr{integral},
\eqr{integral_gen} are valid for an arbitrary representations.

\section{Functional of boundaries and set $\Sigma$.}

Using property \eqr{shift} we proceed further link by link and integrate
out all link variables inside an arbitrary $D$-volume. In this way we
obtain some set $\Sigma$ of (interacting) surfaces $\cS_i$. Suppose, all
of them are {\it disks} (it is clear, at any $D$ it can be done, at least
in a small enough volume). Let $\Gamma_i$ is the boundary of $i$-th disk
and $A_i=A(\cS_i)$ is its (lattice) area. To each $\cS_i$ corresponds the
sum over irreducible representations $r_i$. Corresponding {\it functional
of boundaries} $Z_\Sigma\Big(\{\Gamma\}\Big)$ has the form of a sum over
all configurations $\{r\}$: \be Z_\Sigma\Big(\{\Gamma\}\Big)=\sum_{\{r\}}
\e^{-S_\Sigma(\{r\},\{\d_{\Gamma}\})}
\prod_{i\in\Sigma}d_{r_i}\chi_{r_i}(\Gamma_i)\;,\label{FB1}\ee
\be S_\Sigma\Big(\{r\},\{\d_{\Gamma}\}\Big)=
\ln\sum_{i\in\Sigma}A_iC_2(r_i)-\frac{\l_o}{N}
\sum_{<ij>}A_{ij}\tr\d_{\Gamma_i}\d_{\Gamma_j}\;,\label{SFB}\ee
where $A_{ij}=A(\cS_i\cap\cS_j)$. Formula \eqr{FB1} generalizes the
expression \eqr{fb1} for the functional of boundary in $D=2$. The latter
corresponds to \eqr{FB1} with $\Sigma$ containing only one surface.

There is an infinite number of equivalent sets $\Sigma$. Actually, each
set is defined by the local gauge fixing, i.e., all such a sets are gauge
equivalent. It is clear, observables are independent on the choice
which should be dictated just by convenience of calculations. In $D=3$,
for example, it is possible to chose $\Sigma$ containing surfaces of only
disk topology. The example is drawn (in projection) in Fig.2(a). In this
case, all disks are compressed and has the form of a closed from one side
cylinders of one plaquette width, which fill densely a $3D$ volume.
Choosing another direction of gluing at (at least) one link (see Fig.1)
we obtain another set $\Sigma'$ which differ from $\Sigma$ not only by
smooth deformation of a surfaces but also by an appearance of a compact
surfaces (see Fig.2(b)).

To see what happens when the compact surfaces appear in the set we,
first, consider the situation when the surfaces with a boundaries
are not only disks (1-holed spheres) but also spheres with an arbitrary
number of holes. It is not difficult to see that in such a case we
have to replace an expression $d_{r_i}\chi_{r_i}(\Gamma_i)$ which
appears in \eqr{FB1} and corresponds to $i$-th disk by
$d^{2-n_i}_{r_i}\prod_{k=1}^{n_i}\chi_{r_i}
(\Gamma_{i,k})$ where $\Gamma_{i,k}$ is the boundary of $k$-th hole
($k=1,...,n_i$) on $i$-th surface.

Thus, the most general expression for {\it functional of boundaries} is:
\be Z_\Sigma\Big(\{\Gamma\}\Big)=\sum_{\{r\}}
\e^{-S_\Sigma(\{r\},\{\d_{\Gamma}\})}
\prod_{i\in\Sigma}d^{2-n_i}_{r_i}\prod_{k=1}^{n_i}
\chi_{r_i}(\Gamma_{i,k})\;.\label{FB}\ee
A compactification of a surfaces will become clear in the next section,
where we will consider a partition functions.

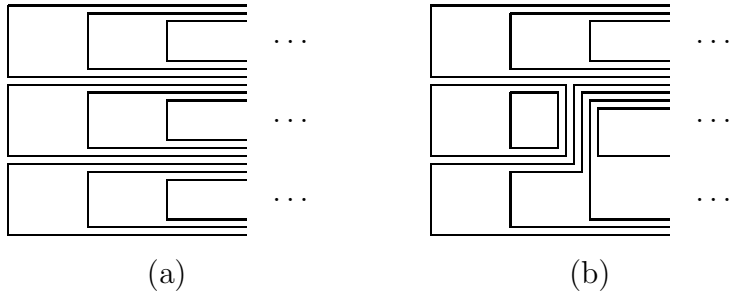
\begin{figure}\centering\begin{picture}(200,100)(- 100,- 25)
\thinlines
\put(-140, 0){\line(1,0){90}}\put(-140,27){\line(1,0){90}}
\put(-140,30){\line(1,0){90}}\put(-140,57){\line(1,0){90}}
\put(-140,60){\line(1,0){90}}\put(-140,87){\line(1,0){90}}
\put(-110, 3){\line(1,0){60}}\put(-110,24){\line(1,0){60}}
\put(-110,33){\line(1,0){60}}\put(-110,54){\line(1,0){60}}
\put(-110,63){\line(1,0){60}}\put(-110,84){\line(1,0){60}}
\put(-80, 6){\line(1,0){30}}\put(-80,21){\line(1,0){30}}
\put(-80,36){\line(1,0){30}}\put(-80,51){\line(1,0){30}}
\put(-80,66){\line(1,0){30}}\put(-80,81){\line(1,0){30}}
\put(-140, 0){\line(0,1){27}}\put(-140,30){\line(0,1){27}}
\put(-140,60){\line(0,1){27}}\put(-110, 3){\line(0,1){21}}
\put(-110,33){\line(0,1){21}}\put(-110,63){\line(0,1){21}}
\put(-80, 6){\line(0,1){15}}\put(-80,36){\line(0,1){15}}
\put(-80,66){\line(0,1){15}}

\put(20, 0){\line(1,0){90}}\put(20,27){\line(1,0){54}}
\put(20,30){\line(1,0){51}}\put(20,57){\line(1,0){51}}
\put(20,60){\line(1,0){90}}\put(20,87){\line(1,0){90}}
\put(50, 3){\line(1,0){60}}\put(50,24){\line(1,0){27}}
\put(50,33){\line(1,0){18}}\put(50,54){\line(1,0){18}}
\put(50,63){\line(1,0){60}}\put(50,84){\line(1,0){60}}
\put(80, 6){\line(1,0){30}}\put(83,30){\line(1,0){27}}
\put(80,51){\line(1,0){30}}\put(80,66){\line(1,0){30}}
\put(80,81){\line(1,0){30}}\put(20, 0){\line(0,1){27}}
\put(20,30){\line(0,1){27}}\put(20,60){\line(0,1){27}}
\put(50, 3){\line(0,1){21}}\put(50,33){\line(0,1){21}}
\put(50,63){\line(0,1){21}}\put(80,66){\line(0,1){15}}
\put(80, 6){\line(0,1){45}}\put(68,33){\line(0,1){21}}
\put(71,30){\line(0,1){27}}\put(74,27){\line(0,1){30}}
\put(77,24){\line(0,1){30}}\put(74,57){\line(1,0){36}}
\put(77,54){\line(1,0){33}}\put(83,30){\line(1,0){27}}
\put(83,48){\line(1,0){27}}\put(83,30){\line(0,1){18}}

\put(-80,-15){\makebox(0,0){(a)}}
\put( 80,-15){\makebox(0,0){(b)}}

\put(-40,13){\dots}\put(-40,43){\dots}\put(-40,73){\dots}
\put(120,13){\dots}\put(120,43){\dots}\put(120,73){\dots}

\end{picture}
\caption[x]{\footnotesize Two examples of gauge equivalent sets of
surfaces (projection from $D=3$). This equally can be considered as a set
of curves for the Principal Chiral Field model.}\label{fig2}\end{figure}

\section{Evolution operator, partition functions and loop averages.}

The expression \eqr{FB} can be equally written as
\be Z_\Sigma\Big(\{\Gamma\}\Big)=\hat{R}\prod_{i\in\Sigma}Z(\Gamma_i)\;,
\label{FB2}\ee \be\hat{R}=\exp\frac{\l_o}{N}\sum_{<ij>}A_{ij}\tr
\d_{\Gamma_i}\d_{\Gamma_j}\;,\label{R}\ee
where the $2D$ boundary functional $Z(\Gamma_i)$ is given by \eqr{fb1} in
the case of a disk or by \eqr{fb} in a general case (then, $\Gamma_i$
means the set of boundaries $\{\Gamma_{i,1},...,\Gamma_{i,n_i}\}$
corresponding to $i$-th surface). Thus, the differential operator
$\hat{R}$ describes an {\it evolution} of QCD$_2$ in QCD$_D$.

The expression for functional of boundaries \eqr{FB} can serve as the
building block in construction of partition functions and observables on
arbitrary $D$-manifolds including those of non-trivial topology.

It is clear, the partition function can be obtained from \eqr{FB} by
putting all boundaries $\Gamma_i$ be equal to $I$, i.e.,
\be Z_\Sigma=\sum_{\{r\}}\e^{-S_\Sigma(\{r\},\{\d_{\Gamma=I}\})}
\prod_{i\in\Sigma}d^{2-n_i}_{r_i}
\prod_{k=1}^{n_i}\chi_{r_i}(\Gamma_{i,k})\;,\label{Z}\ee
where $S_\Sigma(\{r\},\{\d_{\Gamma}\})$ is defined in \eqr{SFB}. This is
a general expression for the partition function of U(N) (and SU(N))
lattice quantum gauge theory in $D$-dimensions.
\bigskip

To write the Wilson loop average \eqr{W} in the same terms let us
consider an arbitrary surface $\cS_C$ such that $C=\d \cS_C$. Then,
the result is \be W(C)=\frac{1}{NZ_\Sigma}\sum_{\{r\}}
\e^{-S_\Sigma(\{r\},\{\d_{\Gamma=I}\})}
\prod_j\e^{\frac{\l_o}{N}A(\cS_C\cap\cS_j)\tr\d_{C=I}\d_{\Gamma_j=I}}
\tr C\prod_{i\in\Sigma}d^{2-n_i}_{r_i}\prod_{k=1}^{n_i}\chi_{r_i}
(\Gamma_{i,k})\;.\label{WLA}\ee

Differentiation with respect to $C$ is easy to perform and we obtain an
equivalent expression:\be W(C)=\frac{1}{NZ_\Sigma}\sum_{\{r\}}
\e^{-S_\Sigma(\{r\},\{\d_{\Gamma=I}\})}
\tr\Big(\prod_j\e^{\frac{i\l_o}{N}A(\cS_C\cap\cS_j)\d_{\Gamma_j=I}}\Big)
\prod_{i\in\Sigma}d^{2-n_i}_{r_i}\prod_{k=1}^{n_i}\chi_{r_i}
(\Gamma_{i,k})\;.\label{WLA2}\ee
Thus, the loop average takes the form of an average of
differential operator \be\frac{1}{N}\tr\Big(\prod_j
\e^{\frac{i\l_o}{N}A(\cS_C\cap\cS_j)\d_{\Gamma_j}}\Big)\;.\ee
It is not difficult to check that the result is independent on a choice
of $\cS_C$.

\section{Discussion.}

We represented pure gauge lattice QCD$_D$ as a statistical model of
integer-valued scalar fields living on a set of interacting surfaces.
Apparently, modulo some possible (and hopefully fruitful) re-writing, the
expressions derived here for the functional of boundaries \eqr{FB}, for
the partition function \eqr{Z} and for loop averages \eqr{WLA} cannot be
simplified further, unless continuum limit is taken. This can be seen
already in the U(1) case (QED), where partition function takes the simple
form: \be\ba{lll}&\dsp Z_\Sigma=\sum_{\{n\}}\e^{-S_\Sigma(\{n\})}\;,\\&\\
&\dsp S_\Sigma(\{n\})=\frac{\l_o}{2}\sum_{i\in\Sigma}A_in_i^2-
\l_o\sum_{<ij>}A_{ij}n_in_j\;.\ea\label{ZQED}\ee
By ``simplification" we mean an essential reduction of
configuration space $\{r\}$. Such a simplification is expected in
continuum limit. We hope that the derived integrated version of the
model is a better starting point for analytical study the continuum
limit than the original, non-integrated version. The problem might
perhaps be solvable by methods of elementary combinatorics, where
the only subtlety is the calculation of the surface entropy factor
which manifests itself in a change of the set $\Sigma$ under refinement
of the lattice \footnote{I wish to thank M.Karliner for discussion
of this point.}.

The method of integration over unitary matrices and the formula
\eqr{integral} (or, in general, \eqr{integral_gen}) can be
applied without any changes to the {\it Principal Chiral Field} (PCF)
model (for recent developments in the model and for references see
\cite{FKW}). The set $\Sigma$ in this case becomes a set of curves
(see Fig.2) and the partition function is again of the form \eqr{Z},
after we replace areas of a surfaces by lengths of curves. Apparently,
due to its respective simplicity, the PCF model might become the first
case where the continuum limit can be taken.

The problem of integration over unitary matrices in QCD is similar to one
in matrix models of $2d$ quantum gravity embedded in $D>1$ (the simplest
case where the problem appears is the closed $D=1$ target space, i.e.,
closed chain of hermitian matrices). The present method can be
applied, after some modification, to these models as well.

Among possible direct continuations of our analysis let us mention the
following:

\ni(i) Expression \eqr{FB} looks suitable for $1/N$-expansion.
The latter has been recently elaborated in a detail for $D=2$
\cite{Gross}-\cite{GT2} \footnote{For $1/N$ applied to generalized $D=2$
Yang-Mills see \cite{TAU}.}. It is tempting to apply this technique to
$D>2$. A straightforward strategy would be to find out a ``stringy"
interpretation for the evolution operator \eqr{R}. Then the whole theory
could be considered as a set of stringy models (QCD$_2$), interacting
through the operator $\hat{R}$. However, a less naive strategy is
possible, if one makes the $1/N$-expansion of whole expression \eqr{FB},
together with $\hat{R}$-operator. Then, each term of $1/N$-expansion
will take a form of a sum over all surfaces from the set $\Sigma$. This
requires some new technique, especially with respect to
$\hat{R}$-operator.

\ni(ii) The derived exact expression for functional of boundaries
\eqr{FB}, especially written in the form \eqr{FB2}, can serve as a
starting point for a {\it mean field} analysis of the model
\footnote{For a mean field analysis of the model on a {\it tree} see
\cite{Bou}.}. An equation for eigenvalues of $\hat{R}$-operator, under
mean field assumption, seems solvable, at least on a finite lattices.

\ni(iii) It is not difficult to recognize that the expression \eqr{Z}
for QCD partition function can be interpreted at infinite $N$ as the
constrained matrix model. The corresponding technique was worked out
for $D=2$ in \cite{infN}-\cite{RY}. It would be interesting to solve the
large N saddle point equation and to see if there is a region of the
coupling constant where unitary constraint can be ignored, which would
indicate an existence of large N (apparently, third order) phase
transition \cite{GW}.

\bigskip

\ni
{\large\bf Acknowledgments.}\\
\ni I wish to thank High Energy Group of Tel-Aviv University for kind
hospitality.

\bb
\bibitem{Wils} K.Wilson, Phys. Rev. {\bf D10} (1974) 2445.
\bibitem{Mig} A.A.Migdal, ZhETF {\bf 69} (1975) 810
                         (Sov. Phys. JETP {\bf 42} 413).
\bibitem{Rus} B.Rusakov, Mod. Phys. Lett. {\bf A5} (1990) 693.
\bibitem{Witten}  E.Witten, Comm. Math. Phys. {\bf 141} (1991) 153;
                            J. Geom. Phys. {\bf 9} (1992) 303.
\bibitem{Zel} D.Zelobenko, {\em Compact Lie groups and their
                          applications}, Moscow, Nauka, 1970
                  (Amer.Math.Soc.Translations {\bf 40}, 1973).
\bibitem{FKW} V.A.Fateev, V.A.Kazakov and P.B.Wiegmann,
                           Nucl. Phys. {\bf B424} (1994) 505.
\bibitem{Gross} D.Gross, Nucl. Phys. {\bf B400} (1993) 161.
\bibitem{GT1}   D.Gross and W.Taylor, Nucl. Phys. {\bf B400} (1993) 181.
\bibitem{GT2} D.Gross and W.Taylor, Nucl. Phys. {\bf B403} (1993) 395.
\bibitem{TAU} O.Ganor, J.Sonnenschein and S.Yankielowicz, {\em The
              string theory approach to generalized 2D Yang-Mills
              theory}, TAUP-2182-94 (July 1994), hep-th/9407114.
\bibitem{Bou} D.Boulatov, {\em QCD on a tree}, NBI-HE-94-24 (April 1994),
                           hep-th/9404136.
\bibitem{infN}  B.Rusakov, Phys. Lett. {\bf B303} (1993) 95.
\bibitem{DK}  M.Douglas and V.Kazakov, Phys. Lett. {\bf B319} (1993) 219.
\bibitem{RY} B.Rusakov and S.Yankielowicz, {\em Large N phase transitions
                    and multi-critical behaviour in generalized 2D QCD},
                    CERN-TH.7390/94, TAUP-2191-94, hep-th/9408046,
                   accepted for publication in Phys. Lett. {\bf B}.
\bibitem{GW}    D.Gross and E.Witten, Phys. Rev. {\bf D21} (1980) 446.
\eb

\vskip1cm

{\LARGE\bf Appendix.}

\bigskip

In this appendix, we will demonstrate on a few examples an action of
operator $\exp\g\tr\d_A\d_B$ ($\g$ is a parameter) on the function
$\chi_{r_1}(A)\chi_{r_2}(B)$. Thus, we calculate
\be\e^{\g\tr\d_A\d_B}\chi_{r_1}(A)\chi_{r_2}(B)\;.\label{ap1}\ee

(a) If $r_1$ or $r_2$ is trivial representation $\{1,0,...,0\}$, then
\eqr{ap1} is equal to 1.

(b) The first non-trivial example is $r_1=r_2=\{1,0,...,0\}$
(fundamental representation). In this case we have
\be\e^{\g\tr\d_A\d_B}\tr A~\tr B=\sum_{q=0}^\infty\gq\tr^q\d_A\d_B
\tr A~\tr B=\tr A~\tr B~\cosh\g-\tr AB~\sinh\g \ee since the even-order
derivatives result in $\tr A~\tr B$ and odd-order ones result in
$\tr AB$.

(c) $r_1=\{1,0,...,0\}$, $r_2=\{1,1,0,...,0\}$. Then
\be\ba{ll}&\dsp\e^{\g\tr\d_A\d_B}~\tr A~\frac{1}{2} (\tr^2 B-\tr B^2)\\
&\dsp=(\tr B~\tr AB~-\tr AB^2)\frac{\e^{2\g}-\e^{-\g}}{3}
+\tr A(\tr^2 B~-\tr B^2)\frac{\e^{2\g}+2\e^{-\g}}{6}\;.\ea\ee

(d) $r_1=\{1,0,...,0\}$, $r_2=\{2,0,...,0\}$. Then
\be\ba{ll}&\dsp\e^{\g\tr\d_A\d_B}~\tr A~\frac{1}{2} (\tr^2 B+\tr B^2)\\
&\dsp=(\tr B~\tr AB~+\tr AB^2)\frac{\e^{-2\g}-\e^{\g}}{3}
+\tr A(\tr^2 B~+\tr B^2)\frac{\e^{-2\g}+2\e^{\g}}{6}\;.\ea\ee

We leave it to reader to substitute these results in the formula
\eqr{integral} to see how it works in these cases.

\end{document}